\documentstyle[prb,aps,twocolumn,epsfig,amsmath]{revtex}

\begin{document}
\twocolumn[\hsize\textwidth\columnwidth\hsize\csname @twocolumnfalse\endcsname
\title{Canonically conjugate pairs and phase operators}

\draft
\author {K.Sch\"onhammer}
\address{Institut f\"ur Theoretische Physik, Universit\"at
  G\"ottingen, Bunsenstr. 9, D-37073 G\"ottingen, Germany}

\date{\today}
\maketitle

\begin{abstract}

For
quantum mechanics on a lattice the position (``particle number'')
operator and the quasi-momentum (``phase'') operator
 obey canonical
commutation relations (CCR) only on a {\it dense set} of the Hilbert space.
We compare exact numerical results for a particle in simple
 potentials on the lattice with the expectations, when 
the CCR are assumed to be {\it strictly} obeyed. Only for sufficiently smooth
eigenfunctions this leads to reasonable results. In the long time
limit the use of the CCR can lead to a {\it qualitatively} wrong
dynamics even if the initial state is in the dense set.
  
\end{abstract}

\vskip 2pc]
\vskip 0.1 truein
\narrowtext

Since the pioneering work on quantization of the electromagnetic
field, where Dirac \cite {Dirac} introduced a phase observable,
supposedly conjugate to the number operator $\cal N$,
 there has been a long
controversy wether a phase operator $\hat \theta$ can be constructed
 which obeys the canonical commutation relation (CCR) $[{\cal N},\hat
 \theta]=i \hat 1$.
 As there exist excellent reviews which address this question
\cite {CN,R2} it is not necessary to give a detailed history of the
various proposals. We want to point out that there are different
types of ``number operators'' of importance.
In quantum optics the eigenvalues of the particle number operator run
from {\it zero} to infinity. For the case of fermions with a
 filled Dirac sea
one often works with the number operator relative to the Dirac sea
which has integer eigenvalues which run from minus infinity to plus
infinity \cite {H,DS,S}.
 A conceptionally simple case where the number operator is of the
form ${\cal N}= \sum^\infty_{-\infty}|n \rangle n \langle n|   $
is (apart from the multiplicative factor of the lattice spacing $a$)
the position operator $\hat x$ for quantum mechanics on a one-dimensional
{\it  lattice} \cite {B}.
In this paper case we treat infinite lattices
which corresponds to the simpler
case of a number operator {\it not} bounded from below \cite {R2}, 
while the quantum optics scenario corresponds to a semi-infinite lattice.

Shortly after Dirac's introduction of the phase operator Jordan \cite {Jordan}
gave an argument why the CCR $[{\cal N},\hat
 \theta]=i \hat 1$ cannot hold as an operator identity. This commutation
relation would imply $[\sin{\alpha \cal N},\hat \theta]
=i\alpha \cos{\alpha \cal N }$. If one puts $\alpha=\pi$ this leads to
a {\it contradiction} as  $\sin{\pi \cal N}$ is zero because $\cal N $
has integer eigenvalues, but  $\cos{\pi \cal N} $ is nonzero.
It is only possible to construct a canonical pair
 $ ({\cal N},\hat \theta)$ 
on a {\it dense  set} of the states of the Hilbert space
\cite{GW,G}. 
 The question we want to address in this paper is if this
fact is useful for typical applications in physics, where a
Hamiltonian is given and one wants to solve the time-independent 
and (or) the time-dependent Schr\"odinger equation.
An example where in this context the canonical commutation relation
 $[{\cal N},\hat  \theta]=i \hat 1$ is generally assumed is
in treatments of the Josephson Hamiltonian \cite {AL}.

As a simple example we study the
properties of a particle described by the Hamiltonian
\begin{equation}
H=-\sum_{m,n}\left (|m+n\rangle t_n \langle m|+H.c.\right )
 +\sum_m|m\rangle V_m
\langle m|,
\end{equation}
where the $t_n$ are the hopping matrix elements and the $V_m$
are the local potentials. 
For a constant potential the (improper) eigenstates are the Bloch
states 
\begin{equation}
|k\rangle =\sqrt{\frac{a}{2\pi}}\sum_me^{iakm}|m\rangle,
\end{equation} 
with $k$ in the first Brillouin zone
$ (-\pi/a,\pi/a]$. The Bloch states are
 normalized as $\langle k|k'\rangle=\delta(k-k')$. The ``obvious''
operator to replace the momentum operator (divided by $\hbar$) on the
lattice is the quasi-momentum operator
\begin{equation}
\hat k=\int^{\pi/a}_{-\pi/a}|k\rangle k \langle k|dk.
\end{equation}
In the position representation 
$\hat \theta \equiv a\hat k$ is just  
the infinite lattice version of the ``phase operator'' proposed
independently by Garrison and Wong \cite {GW} and Galindo \cite {G}
for the semi-infinite case
\begin{equation}
\hat \theta |n\rangle =\sum_{m (\ne n)}\frac{(-1)^{m-n}}{i(m-n)}
|m\rangle.
\end{equation}
If one applies $\hat x = a\cal N$  to the states in Eq. (4) one sees that
$ \hat x \hat \theta |n\rangle $ is not an element of
the Hilbert space $l_2$,
while $  \hat \theta \hat x |n\rangle $ is.
A straightforward formal calculation shows that 
$\hat x $ and $\hat k$ do {\it not} obey a CCR
\begin{equation}
\left [ \hat x,\hat k \right ]|n\rangle =
-i\sum_{m (\ne n)}(-1)^{m-n}|m\rangle.
\end{equation}
The correction term can be expressed in terms of the projection
operator on the Bloch state $|\pi/a\rangle$ \cite {Judge}
\begin{equation}
\left [ \hat x,\hat k \right ]
=i\left (\hat 1 -\frac{2\pi}{a}\left |\frac{\pi}{a}
\rangle\langle \frac{\pi}{a}\right |\right ).
\end{equation}
For states $|\psi\rangle $ where the components $\psi_n\equiv \langle
 n|\psi \rangle$ obey the condition 
$S_\psi\equiv \sum_n (-1)^n \psi_n=0$ the correction term
vanishes. These states form a {\it dense set} of states of the 
Hilbert space $l_2$ \cite {G}. This condition is fulfilled
when the components in position space are sufficiently smooth.
If $\psi_n=f(an)$, where $f$ is a differentiable function,
the operator $i\hat k$ for $a \to 0$ acts like the differential
operator $d/d(an)$. Using Eq. (4) this can be seen by writing
\begin{equation}
\langle n|i\hat k|\psi\rangle
=\sum^\infty_{j=1}\frac{(-1)^{j+1}}{ja}\left ( \psi_{n+j}- \psi_{n-j}
\right ).
\end{equation}
With the assumption on the function $f$ made above the series on the
right hand side (rhs) is approximately  
of the form $2(f'(an)-f'(an)+ f'(an)-+...)$
which quickly converges to $f'(an) $. This can be best seen using the 
Euler transformation \cite {AS} of the sum on the rhs of Eq. (7).

We next consider {\it functions} of the operator $\hat k$. For integer
$n$ the exponential functions
\begin{equation}
e^{-ina\hat k}=\sum_m |m+n\rangle\langle m|\equiv \hat T_n
\end{equation}
are just the discrete translation operators. Using the site
representation one easily shows the commutation relation
$[\hat x,e^{-ina\hat k} ]=ane^{-ina\hat k}$, which also can be
``derived'' assuming the CCR. The latter incorrect derivation
could also be performed
 if $n$ is replaced by an arbitrary real number. But then
 the above result no longer holds, i.e. the use of the CCR gives
a wrong answer. 
%The fact that the CCR ``derivation'' gives
%the correct answer for integer $n$ is the main reason why
%the CCR assumption can sometimes lead to correct results. 
In the following we also use the operator $\hat k^2$. Its matrix
elements in the position representation are given by
\begin{equation}
a^2\langle m|\hat k^2|n\rangle=\frac{\pi^2}{3}\delta_{mn}
+2\frac{(-1)^{m-n}}{(m-n)^2}(1-\delta_{mn}). 
\end{equation}
Therefore the operator $\hat k^2/2$ 
has the form of the first term on the rhs of Eq. (1)
with $t_0=-\pi^2/(6a^2)$ and $t_n=(-1)^{n+1}/(an)^2$. The corresponding
energy eigenvalues $\epsilon_k=-t_0-2\sum_{n\ge 1}t_n\cos{(akn)}$
are just the well known Fourier series which corresponds to
 the periodically
 continued
parabola arcs. In the following we mainly work with
the Hamiltonian $H=\hat k^2/2+V(\hat x)$, which is a special case 
of Eq. (1). 
 
 We examine, if {\it assuming} the strict
validity of the CCR $[\hat x ,\hat k]=i\hat 1$ can be useful
in the description of a particle governed by such a Hamiltonian
by comparison with exact results.
We study the eigenvalue problem as well as solutions to the
time dependent Schr\"odinger equation. 
  Only
for those eigenstates $|\psi_E \rangle $ which approximately fulfill
$\langle \pi/a|\psi_E\rangle=0$  we can expect the assumption 
 $[\hat x, \hat k]=i\hat 1$ to be useful.

We begin with the {\it harmonic oscillator on the lattice}
$H=\hat k^2/2+c\hat x^2/2$.
The Josephson Hamiltonian in its most simple form \cite {AL} correponds
to a quadratic potential but with a 
 kinetic energy $\hat T_{kin}$ with
 nearest neighbour hopping $(\hat 1-\cos{a\hat k})/a^2
=[\hat 1 -(T_1+T_{-1})/2]/a^2$ for $a=1$.
We present results for both types of the kinetic energy
and explicitely mention when $\hat T_{kin}=\hat k^2/2$ is {\it not} used .
 The exact solution can be obtained
numerically by diagonalizing a sufficiently large matrix. 
The expectations from assuming the CCR $[\hat x ,\hat k]=i\hat 1$  
are obvious if one introduces the usual ladder operators \cite {B}
and determines the eigenvalue spectrum purely algebraically,
which yields $E_n=\sqrt{c}(n+1/2)$. If we further assume the
expectation value of the potential energy to be half the energy
eigenvalue as in the continuum limit it follows that the 
the wave function $\psi^{(n)}_m$ 
approximately extends from $-(n/\sqrt{c})^{1/2}$ to  $(n/\sqrt{c})^{1/2}$.
The mean separation of the nodes of the corresponding wave function in 
the continuum is therefore given by $2/\sqrt{nc^{1/2}}$. The smoothness     
assumption which is necessary for the CCR to be approximately valid
certainly breaks down, when this separation becomes of the order of the 
lattice constant $a$. One therefore expects the equidistant energy
spectrum of the harmonic oscillator in the continuum to be
approximately valid for 
quantum numbers $n$ smaller than $b/\sqrt{a^4c}$
where $b$ is of order unity.
The exact numerical results for the lattice model are shown in Fig. 1.

\begin{figure}[hbt]
\begin{center}
\epsfig{file=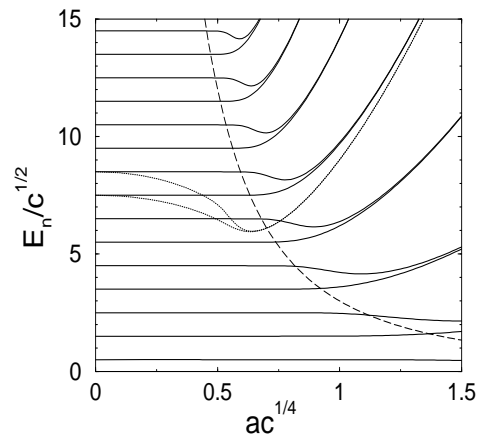,width=6cm,height=6cm,angle=-90}
\end{center}
\caption{Normalized  energy eigenvalues of the harmonic oscillator on
 the lattice as a function of  $ac^{1/4}$ .
 Below the dashed line $3/(ac^{1/4})^2$ explained in the text
one approximately recovers the eigenvalue spectrum of the continuum
limit.
The dotted line shows two eigenvalues for the kinetic energy
$(\hat 1-\cos{a\hat k})/a^2$.   }
\end{figure}

\noindent The dashed line shows the criterion just
derived for $b=3$. Above this line there are strong deviations from
the equidistant spectrum, in fact pairs of eigenvalues
 become almost degenerate.
A simple explanation for this observation will be given later.
For comparison we also show two eigenvalues for the nearest
neighbour kinetic energy $(\hat 1-\cos{a\hat k})/a^2$ which for
small enough values of $ac^{1/4}$ are expected to form the same equidistant
spectrum. 
 As shown the results deviate from 
 $E_n=\sqrt{c}(n+1/2)$ already for smaller values of $ac^{1/4}$
and the almost degeneracy of the eigenvalue pair occurs already for
a smaller value of  $ac^{1/4}$ compared to the long range hopping.
 Fig. 2 shows the behaviour of the 
absolute value of the overlap $S_n\equiv \sum_m (-1)\psi^{(n)}_m$
as a function of the quantum number $n$ for the even parity states.      
The values of $n$ where $S_n$ starts to become ``nonzero''
 agree very well with expectation from the argument using the number
of nodes. 

In order to understand the spectrum in the upper right corner of
Fig. 1 it is useful to first treat the case of a {\it linear} potential
described by the Hamiltonian $H=\hat k^2/2-F\hat x$.
 For such a
potential the energy eigenvalue spectrum for the lattice model
is known to be completely different from the continuum limit. It
consists of an equidistant ``Wannier-Stark''-ladder \cite {WS}, while
the spectrum is purely continous for the continuum
case. This can easily 
be understood using the translation
 operators $\hat T_n $ defined in Eq. (8)
 which obey the simple commutation relation
$[\hat T_n,H]=anF\hat T_n$. After one localized eigenstate 
$|E_0\rangle $ with eigenvalue $E_0$ is shown 
to exist \cite {WS} this commutation relation implies that
the $\hat T_n
|E_0\rangle$ are  eigenstates with eigenvalues $E_0-anF$. The form of 
the localized Wannier-Stark states for the long range hopping of
Eq. (9) is shown in Fig. 3 (stars).
 The other wave function (circles) is discussed when we return to the
discussion of the harmonic potential.

\begin{figure}[hbt]
\begin{center}
\epsfig{file=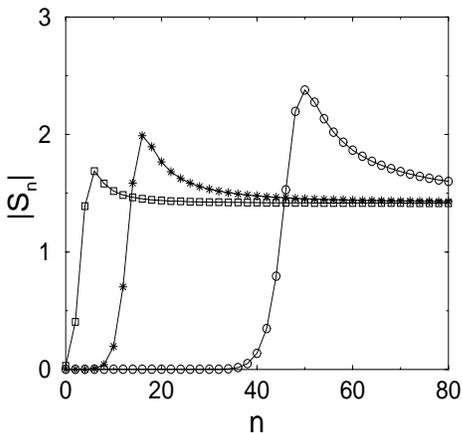,width=6cm,height=6cm,angle=-90}
\end{center}
\caption{Absolute value of the overlap   $S_n\equiv \sum_m
  (-1)\psi^{(n)}_m$ which determines the deviation from the CCR
as a function of the quantum number $n$ for the even parity states,
and different values of $c$: $c=1$ squares, $c=0.1$ stars and
$c=0.01$ circles. }
\end{figure}

% While assuming the validity of the CCR was of restricted use for the
% understanding of the spectrum of the harmonic potential this is not
% the case here.
 We next examine if the CCR assumption is of any use
for the understanding of the {\it time evolution} in the linear
potential.
The equations of motion for the Heisenberg operators $\hat x(t)\equiv
e^{iHt}\hat x e^{-iHt}$ and $\hat k(t)$ close if the CCR are assumed and yield
$\hat k(t)_{\rm CCR}=\hat k +Ft\hat 1$ as well as
$\hat x(t)_{\rm CCR}=\hat x +\hat k t +\hat 1Ft^2/2$.
Obviously the result for $\hat k(t)_{\rm CCR}$ is incorrect
as the operator $\hat k$ defined in Eq. (3) is {\it bounded}
and this property also has to hold for $\hat k(t)$ .
The result for the position operator using the CCR 
differs from the {\it exact} solution, which follows from 
$[\hat x,\hat T_n]=an\hat T_n$ and $[\hat T_n,H]=anF\hat T_n$ as
\begin{equation}
\hat x(t)= \hat x +\sum_{n>0}\left [ t_n\hat
  T_n\left(e^{-ianFt}-1\right )/F+H.c.\right ].
\end{equation}
This shows that the exact solution is {\it periodic } in time. The
particle undergoes so called ``Bloch-oscillations''\cite {WS}.
 \begin{figure}[hbt]
\begin{center}
\epsfig{file=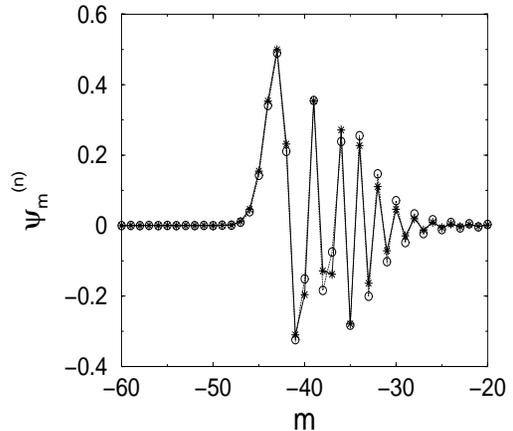,width=6cm,height=6cm,angle=-90}
\end{center}
\caption{The stars show the amplitudes 
$\sqrt{2} \psi_m$ of the Wannier-Stark eigenstate for a linear
  potential on a lattice with $F=0.4$ localized around the site
$m=-41$. The circles show the amplitudes  $\psi_m$ of the even eigenstate
of the harmonic potential with $c=0.01$ localized around the same site.}
\end{figure}

The two results are compared in Fig. 4 for two Gaussian wave pakets
($\psi_n(t=0)\sim \exp{[-b(n-n_0)^2}]$) 
with the same location but different widths as
initial states.
 \begin{figure}[hbt]
\begin{center}
\epsfig{file=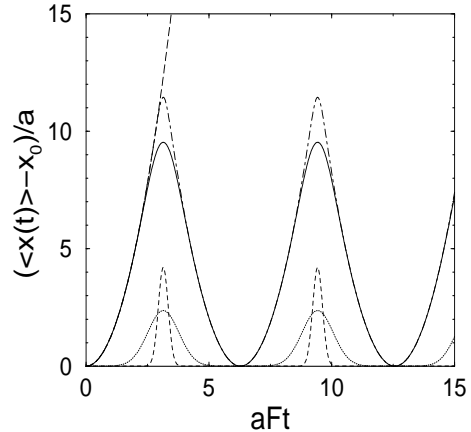,width=6cm,height=6cm,angle=-90}
\end{center}
\caption{The full curve ($b=0.2$) and the dashed-dotted ($b=0.02$)
curve show the
periodic Bloch
oscillations in a linear potential ($F=0.4$). The long dashed curve
shows the result from assuming the CCR. In order to understand
why this approximation fails qualitatively when approaching the time
$\pi/(aF)$  the time dependent overlaps $|S_{\psi(t)}|$ 
(see text) are shown as the
dotted  ($b=0.2$) and dashed curves ($b=0.02$) .}
\end{figure}
\noindent If the initial wave paket extends over many lattice
sites the CCR assumption yields an excellent approximation for times
less than $\pi/(aF)$. At this time the turning point is reached in the
exact solution. 
Approaching this time the overlap $|S_{\psi(t)}|
\equiv |\sum_n(-1)\psi_n(t)|$,
also shown in Fig. 4,
becomes large and the CCR assumption breaks down and yields a {\it
  qualitatitely} wrong long time behaviour with free acceleration as
in the continuum. It should be pointed out that
for a {\it linear potential} using a
$2\pi$-periodic
funtion of $a\hat k $ for the kinetic energy, like
e.g.  the nearest neighbour hopping form $(\hat 1-\cos{a\hat k})/a^2$
the CCR result for  $\hat x(t)$ accidentally 
agrees with the exact solution.
This is related to the fact mentioned earlier that the
wrong use of the CCR yields the exact commutation realation
 $[\hat x,e^{-ina\hat k} ]=ane^{-ina\hat k}$ for integer $n$.

After this discussion of the linear potential we can return to 
the harmonic potential and explain its properties
in the regime where the CCR
assumption breaks down.
\begin{figure}[hbt]
\begin{center}
\epsfig{file=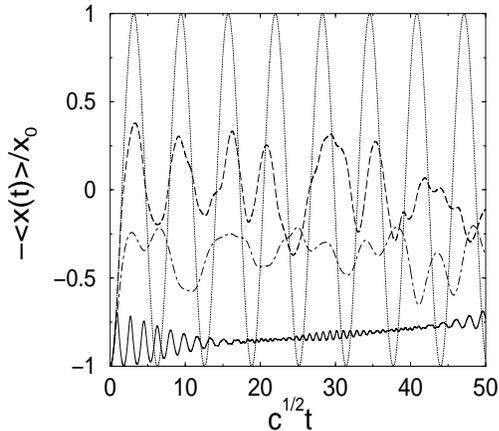,width=6cm,height=6cm,angle=-90}
\end{center}
\caption{Average position of initially Gaussian wave packets (b=0.2) centered
at $x_0=-n_0a$ in a harmonic
  potential ($c=0.01$)
as a funtion of $\sqrt{c}t$. For $n_0=20$ (dotted curve)
the motion is harmonic as predicted by the CCR. For $n_0=30$
(dashed curve) many
different excitation energies contribute and the motion is no longer
harmonic. For $n_0=40$ 
(full curve) the time scale is too short to show the slow
oscillation towards $n_0a$ and back. The dashed-dotted curve
is for the case of nearest neighbour hopping and $n_0=20$.}
\end{figure}
\noindent  The almost degeneracy of pairs of eigenvalues
 shown in the upper right corner of Fig. 1 can now easily be explained.
As shown in Fig. 3 (circles) the eigenstates for large enough quantum
numbers are just even and odd states which (apart from a factor
$\sqrt{2}$) almost look like the Wannier-Stark state which one obtains
when the harmonic potential is linearized locally. The
bonding-antibonding splitting decreases exponentially with the distance
of the the right and left part of the eigenfunctions. This leads to 
a very slow component in the time evolution of the position operator
shown in Fig. 5.
 For the same potential  the time evolution strongly depends on
the initial position $x_0=-n_0a$ of the wave paket.
 For sufficiently small values of $n_0$ the expectation value 
$\langle \hat x (t)\rangle$ is to a very good approximation 
given by a cosine function as predicted by
assuming the CCR. For $n_0=30$ many different
frequencies contribute and lead to clear deviations from
harmonic motion (dashed
curve).
For $n_0=40$ the initial state has strong overlap
with eigenstates for which the bonding-antibonding splitting is
already rather small and the slow oscillation towards $an_0$ cannot be seen
in the figure.
Only in the very short time limit the CCR result for expectation value is
useful. The fact that the deviations get more pronounced by {\it
  increasing}
$n_0$ is opposite to what is widely believed \cite {AL}
but is obvious from Fig.1. 
One obtains qualitatively similar results if the long range hopping
is replaced by nearest neighbour hopping, i.e. $\hat k^2/2 \to (\hat 1
-\cos{a\hat k})/a^2$ (see dashed-dotted curve in Fig. 5), 
which corresponds to the Josephson Hamiltonian.
The two regimes indicated in Fig. 1 correspond to the Josephson and
the Fock regime for this Hamiltonian \cite {AL}. In the latter
regime assuming the CCR must certainly fail for the long time dynamics.

In this paper we have shown that the fact that a canonical
operator which is conjugate to the position operator on the lattice
can be defined on a {\it dense set} of the Hilbert space is of very
restricted use in practical applications. This is an indication that
one has to be extremely careful
in similar use of the particle number-phase CCR in 
 fields like interacting electrons in one dimension \cite {H,DS,S}
and the Josephson effect\cite {AL}.

The author wants to thank V. Meden and W. Zwerger for stimulating
 discussions.


\begin{thebibliography}{*}
\bibitem{Dirac} P.A. M. Dirac, Proc. Roy. Soc. (London),
 {\bf A114}, 243 (1927)
\bibitem{CN} P. Caruthers and N. M. Nieto, Rev. Mod. Phys. {\bf 40},
  411 (1968)
\bibitem{R2} R. Lynch, Physics Reports {\bf 256}, 367 (1995)
\bibitem{H} D. Haldane, J. Phys. C {\bf 14}, 2584 (1981)
\bibitem{DS} J.v. Delft and H. Schoeller, Ann. Phys. (Leipzig){\bf 7},
 225 (1998)
\bibitem{S} K. Sch\"onhammer, Phys. Rev. B{\bf 63}, 245102 (2001)
\bibitem{B} G. Baym {\it Lectures on Quantum Mechanics}
  (Benjamin/Cummings, Menlo Park, 1969)
\bibitem{Jordan} P. Jordan, Z. Phys. {\bf 44}, (1927) 1 .
As this paper is in German it is apparently not well known.
\bibitem{GW} J. C. Garrison and J. Wong, J. Math. Phys. {\bf 11},
2242 (1970)
\bibitem{G} A. Galindo, Lett. in Math. Phys. {\bf 8}, 495 (1984)
\bibitem{AL} For a review which contains many additional references
see: A. Legget, Rev. Mod. Phys. {\bf 73}, 307 (2001)
\bibitem{Judge} A similar commutation relation was derived earlier
for a particle on a ring: D. Judge, Phys. Lett. {\bf 5}, (1962) 189 
\bibitem{AS} M. Abramowitz and I. Stegun, {\it Handbook of
Mathematical Functions} (Dover, New York, 1964)
\bibitem{WS} G. H. Wannier, Phys. Rev. {\bf 117} , 1366 (1960);
a more recent review is given in J. B. Krieger and G. J. Iafrate,
Phys. Rev. B{\bf 33}, 5494 (1986)  
\end{thebibliography}
\end{document}